\documentclass[%
 reprint,
 amsmath,amssymb,
 aps,
 superscriptaddress,
]{revtex4-1}

\usepackage{graphicx}
\usepackage{dcolumn}
\usepackage{bm}
\usepackage{braket}
\usepackage{color}
\usepackage{latexsym}
\usepackage{amsmath,amsthm}
\usepackage{amsfonts}
\usepackage{amsmath,amssymb,amsthm,mathrsfs,amsfonts,dsfont}
\usepackage{subfigure, epsfig}
\usepackage{braket}
\usepackage{ulem}
\usepackage{bm}
\usepackage{enumerate}
\usepackage{color}
\usepackage{graphicx}

\begin{document}

\preprint{APS/123-QED}

\title{Quantum annealing with symmetric subspaces}



\author{Takashi Imoto}
\affiliation{%
 Research Center for Emerging Computing Technologies, National Institute of Advanced Industrial Science and Technology (AIST),
1-1-1 Umezono, Tsukuba, Ibaraki 305-8568, Japan.
}%
\author{Yuya Seki}%
\affiliation{
Graduate School of Science and Technology, Keio University, Hiyoshi 3-14-1, Kohoku-ku, Yokohama 223-8522, Japan
}

 \author{Yuichiro Matsuzaki}
 \email{matsuzaki.yuichiro@aist.go.jp}
\affiliation{%
 Research Center for Emerging Computing Technologies, National Institute of Advanced Industrial Science and Technology (AIST),
1-1-1 Umezono, Tsukuba, Ibaraki 305-8568, Japan.
}%
\affiliation{NEC-AIST Quantum Technology Cooperative Research Laboratory, National Institute of Advanced Industrial Science and Technology (AIST), Tsukuba, Ibaraki 305-8568, Japan}

\date{\today}

\begin{abstract}
Quantum annealing (QA) is a promising approach for not only solving combinatorial optimization problems but also simulating quantum many-body systems such as those in condensed matter physics.
However, non-adiabatic transitions constitute a key challenge in QA.
The choice of the drive Hamiltonian is known to affect the performance of QA because of the possible suppression of non-adiabatic transitions. Here, we propose the use of a drive Hamiltonian that preserves the symmetry of the problem Hamiltonian for more efficient QA. Owing to our choice of the drive Hamiltonian, the solution is searched in an appropriate symmetric subspace during QA. As non-adiabatic transitions occur only inside the specific subspace, our approach can potentially suppress unwanted non-adiabatic transitions. To evaluate the performance of our scheme, we employ the XY model as the drive Hamiltonian in order to find the ground state of problem Hamiltonians that commute with the total magnetization along the $z$ axis. We find that our scheme outperforms the conventional scheme in terms of the fidelity between the target
ground state and the states after QA.

\end{abstract}

\maketitle

\section{Introduction}

Quantum annealing (QA) is a technique for solving combinational optimization problems \cite{kadowaki1998quantum, farhi2000quantum, farhi2001quantum}.
The solution of the problem is mapped to the ground state of the Ising
Hamiltonian, and it is searched in a large Hilbert space
during QA. The adiabatic theorem guarantees that the solution can be found as long as the dynamics is adiabatic.
Many other applications have been proposed for finding the ground state of the Ising Hamiltonian using QA.
One such application is prime factorization \cite{jiang2018quantum, dridi2017prime}.
The shortest-vector problem is a promising candidate for post quantum cryptography, and an approach for solving this problem using QA has been proposed \cite{joseph2020not, joseph2021two}.
Another application of QA is machine learning \cite{adachi2015application, wilson2021quantum, li2019improved, sasdelli2021quantum, date2020adiabatic, neven2008training, neven2012qboost, willsch2020support, winci2020path}.
In addition, an application to clustering has been reported \cite{kumar2018quantum, kurihara2014quantum}.
Further, a method of using QA to perform calculations for topological data analysis (TDA) has been investigated \cite{berwald2018computing}.
Moreover, a method for mapping the graph coloring problem to the Ising model for the QA format has also been proposed \cite{kudo2018constrained, kudo2020localization}.


Several applications of QA to the simulation of condensed matter have also been reported.
For example, simulation of the $\mathbb{Z}_{2}$ spin liquid has been investigated using QA \cite{zhou2021experimental}.
Further, the Kosterlitz--Thouless (KT) phase transition has been experimentally observed \cite{king2018observation}.
In addition, the Shastry--Sutherland Ising Model has been simulated \cite{kairys2020simulating}.
Moreover, Harris et al. investigated the spin-glass phase transition using QA \cite{harris2018phase}.

However, it is difficult to prepare the exact ground state using QA with the existing technology.
The main causes of deterioration are non-adiabatic transitions and decoherence.
In the case of QA for Ising models to solve combinatorial optimization problems, the probability of successfully obtaining the ground state increases with the number of trials.
For a given density matrix $\rho =\sum _n p_n |E_n\rangle \langle E_n|$, where  $|E_n\rangle $ denotes the eigenvector of the problem Hamiltonian and $p_n$ denotes the population,
single-shot measurements in the computational basis randomly select $E_n$
and let us know this energy $E_n$. Hence, as long as the ground state has a finite  population, we can obtain the ground state energy if we perform a large number of measurements.
Meanwhile, if we consider a Hamiltonian having non-diagonal matrix elements with the computational basis, the measurements in the computational basis do not let us know the energy of the Hamiltonian. Instead, we need to measure an observable of $H$ with Pauli measurements. In this case, we obtain $\langle H \rangle = \sum _n p_n E_n$ for a given density matrix $\rho =\sum _n p_n |E_n\rangle \langle E_n|$ in the limit of a large number of measurements, and this is certainly different from the true ground state if the density matrix has a finite population of any excited state.
Therefore, the imperfection of the ground-state preparation induces a crucial error in the estimation of the energy of the ground state of the Hamiltonian having non-diagonal elements. The objective of this study is to address this problem.
Furthermore, there have been reported cases where quantum annealing fails due to the existence of symmetry\cite{francis2022determining, imoto2021preparing}.

In this paper, we propose the use of a drive Hamiltonian that preserves the symmetry of the problem Hamiltonian for more efficient QA, which can potentially suppress non-adiabatic transitions.
The relevant Hamiltonians in condensed matter physics occasionally commute with the total magnetization $S_z=\sum_{j}\sigma_{j}^{(z)}$, and we aim to obtain the ground state of such a problem Hamiltonian. We employ the XY model (flip-flop interaction) as the drive Hamiltonian. As both the problem and the drive Hamiltonian commute with the
total magnetization, the unitary evolution
is restricted to the symmetric subspace with a specific magnetization.
As the total Hamiltonian commutes with the total magnetization, the non-adiabatic transitions are restricted to the same sector with respect to the total magnetization. Thus, the non-adiabatic transitions can potentially be suppressed.
Through numerical simulations, we demonstrate that our method can suppress non-adiabatic transitions for some problem Hamiltonians.

The remainder of this paper is organized as follows. Section \ref{sec:QA} reviews QA.
Section \ref{sec:method} describes our method of using the flip-flop interaction for the drive Hamiltonian. Section \ref{sec:numerical_result} describes numerical simulations 
conducted to evaluate the performance of our scheme, where the deformed spin star and a random XXZ spin chain are selected as the problem Hamiltonians. Finally, Section \ref{sec:conclusion} concludes the paper.

\section{Quantum annealing 
}\label{sec:QA}

In this section, we review QA \cite{kadowaki1998quantum,farhi2000quantum,farhi2001quantum}.
QA was proposed by Kadowaki and Nishimori as a method for solving combinatorial optimization problems \cite{kadowaki1998quantum}.
We choose the magnetic transverse field
\begin{align}
    H_{D}=-B\sum_{j=1}^{L}\sigma_{j}^{(x)}\label{eq:transverse_field}
\end{align}
as the drive Hamiltonian, where $B$ denotes the amplitude of the magnetic field.
In the original idea of QA \cite{kadowaki1998quantum}, the problem Hamiltonian was
chosen as an Ising model
mapped from a combinatorial optimization problem to be solved.
We set the ground state of this Hamiltonian as
\begin{align}
    \ket{\Phi(0)}=\ket{+\cdots+},
\end{align}
where the state $\ket{+}$ denotes the eigenstate of $\sigma^{(x)}$ with an eigenvalue of $+1$.
The total Hamiltonian for QA is written as follows:
\begin{align}
    H=\frac{t}{T}H_{P}+\biggl(1-\frac{t}{T}\biggr)H_{D}\label{conventionalqa},
\end{align}
where $T$ is the annealing time, $H_{D}$ is the drive Hamiltonian, and $H_{P}$ is the problem Hamiltonian.
First, we prepare the ground state of the drive Hamiltonian.
Second, the drive Hamiltonian is adiabatically changed into the problem Hamiltonian according to Eq. (\ref{conventionalqa}).
As long as this dynamics is adiabatic, we can obtain the ground state of the problem Hamiltonian, which is guaranteed by the adiabatic theorem \cite{morita2008mathematical}.

Non-adiabatic transitions and decoherence are the main causes of performance degradation of QA.
Considerable effort has been devoted toward suppressing such non-adiabatic transitions and decoherence.
It is
reported that a non-stoquastic Hamiltonian with off-diagonal matrix elements improves the performance of QA for some problem Hamiltonians \cite{seki2012quantum, seki2015quantum,susa2022nonstoquastic}.
Susa et al. suggested that the accuracy of QA can be improved using an inhomogeneous drive Hamiltonian for specific cases \cite{susa2018exponential, susa2018quantum}.
Further, it has been shown that Ramsey interference can be used to suppress non-adiabatic transitions \cite{matsuzaki2021direct}.
Other methods have been reported for suppressing non-adiabatic transitions and decoherence using variational techniques \cite{susa2021variational, matsuura2021variationally,imoto2021quantum}.
Error correction of QA has been investigated for suppressing decoherence \cite{pudenz2014error}.
In addition, the use of decoherence-free subspaces for QA has been proposed \cite{albash2015decoherence, suzuki2020proposal}.
Moreover, spin-lock techniques have been theoretically proposed to use long-lived qubits for QA \cite{chen2011experimental,nakahara2013lectures,  matsuzaki2020quantum}.

Experimental demonstrations of QA have also been reported. D-Wave Systems Inc. realized QA machines composed of thousands of qubits by using superconducting flux qubits \cite{johnson2011quantum}.
Many experimental demonstrations of QA using such devices have been reported, such as machine learning, graph coloring problems, and magnetic material simulation \cite{kudo2018constrained, adachi2015application, hu2019quantum, kudo2020localization, harris2018phase}.

\section{Method}\label{sec:method}

In this section, we describe our proposal of QA with symmetric subspaces.
In general, when a Hamiltonian has symmetry, it commutes with an observable that is a conserved quantity.
In this case, the Hamiltonian is block-diagonalized, and the eigenstates of the Hamiltonian are divided into subspaces characterized by the conserved quantity.
When the drive Hamiltonian and problem Hamiltonian
have the same conserved quantities,
any non-adiabatic transitions during QA
will occur inside the same subspace, which is in stark contrast to the conventional scheme where non-adiabatic transitions can occur in any subspace.
Owing to such a restriction on non-adiabatic transitions, our scheme can potentially suppress non-adiabatic transitions.

We employ the one-dimensional XY model as the drive Hamiltonian.
The one-dimensional XY model is defined by
\begin{align}
    H=g\sum_{j=1}^{L}\bigl(\sigma_{j}^{(+)}\sigma_{j+1}^{(-)}+\sigma_{j}^{(-)}\sigma_{j+1}^{(+)}\bigr),\ \ \sigma_{L+1}^{(k)}=\sigma_{1}^{(k)}\label{eq:xy_ham},
\end{align}
where $g$ denotes the coupling strength, $L$ is the number of sites, $\sigma_{j}^{(\alpha)}(\alpha=x,y,z)$ are the Pauli matrices defined on the $j$-th site, and $\sigma_{j}^{(\pm)}=\sigma_{j}^{(x)}\pm i\sigma_{j}^{(y)}$.
Further, $S_z$ is a conserved quantity of the XY model.
This model can be mapped to the free fermion system by the Jordan--Wigner transformation; hence, the ground-state energy of this Hamiltonian can be easily calculated using a classical computer. Therefore, when we try to prepare the ground state of this Hamiltonian (by using thermalization,  for example), we can experimentally verify how close the state is to the true ground state by measuring the 
expectation value and variance of
\cite{imoto2021improving}.
It is worth mentioning that, although the XY model was employed as the drive Hamiltonian in previous studies \cite{kudo2018constrained, kudo2020localization}, the purpose was to add a proper constraint to solve graph coloring problems. To the best of our knowledge, our study is the first to use the XY model as the drive Hamiltonian for the suppression of non-adiabatic transitions.

We choose a problem Hamiltonian that commutes with $S_z$.
This guarantees that the non-adiabatic transitions can occur only inside the symmetric subspace. 
Some Hamiltonians used in condensed matter physics also commute with $S_z$.

From the Lieb--Schultz--Mattis theorem \cite{lieb1961two}, it is known that the energy gap between the ground state and the first excited state
of the XY model is $O(1/L)$ if we fix the value of $g$.
Thus, as the number of qubits increases, this energy gap becomes smaller, which can make it difficult to realize adiabatic dynamics. However, as we are going to optimize
the amplitude of the drive Hamiltonian (as explained below), the problem of the small energy gap can be solved by setting an appropriate value of $g$. 

\section{Numerical results}\label{sec:numerical_result}

This section describes numerical simulations performed to compare our scheme
with the conventional scheme.
We employ the Gorini--Kossakowski--Sudarshan--Lindblad (GKSL) master equation for the numerical simulations to account for the decoherence.
For the drive Hamiltonian,
the transverse magnetic field Hamiltonian
described in Eq. (\ref{eq:transverse_field}) is used
in the conventional scheme, while we employ the one-dimensional XY model in Eq. (\ref{eq:xy_ham}).
Moreover, we choose a deformed spin star model and a random XXZ spin chain as the problem Hamiltonians.
We define the estimation error as $|E^{(\rm{true})}_g-E^{(\rm{QA})}_g|$, where $E^{(\rm{true})}_g$ denotes the true ground state energy and $E^{(\rm{QA})}_g$ denotes the energy obtained by QA.
For fair comparison, we optimize the values of $B$ and g to minimize the expectation value of the problem Hamiltonian after QA (or equivalently, to minimize the estimation error).


We introduce the GKSL master equation to consider the decoherence \cite{gorini1976completely, lindblad1976generators}.
The GKSL master equation is given by

\begin{align}
    \frac{d\rho(t)}{dt}=-i[H(t), \rho(t)]+\sum_{n}\gamma[\sigma^{(k)}_{n}\rho(t)\sigma^{(k)}_{n}-\rho(t)],
\end{align}
where $\sigma^{(k)}_{j}(k=x,y,z)$ denotes the Lindblad operators acting on site $j$, $\gamma$ denotes the decoherence rate, and $\rho(t)$ is the density matrix of the quantum state at time $t$.
We solve the GKSL master equation using Qutip \cite{johansson184nation,johansson2012qutip}.
Throughout this paper, we choose $\sigma^{(z)}$ as the Lindblad operator.

\subsection{Deformed spin star model}
\begin{figure}[htbp]
\begin{center}
\begin{minipage}{0.48\textwidth}
\begin{center}
\begin{minipage}{0.90\textwidth}
\begin{center}
\includegraphics[width=0.95\textwidth]{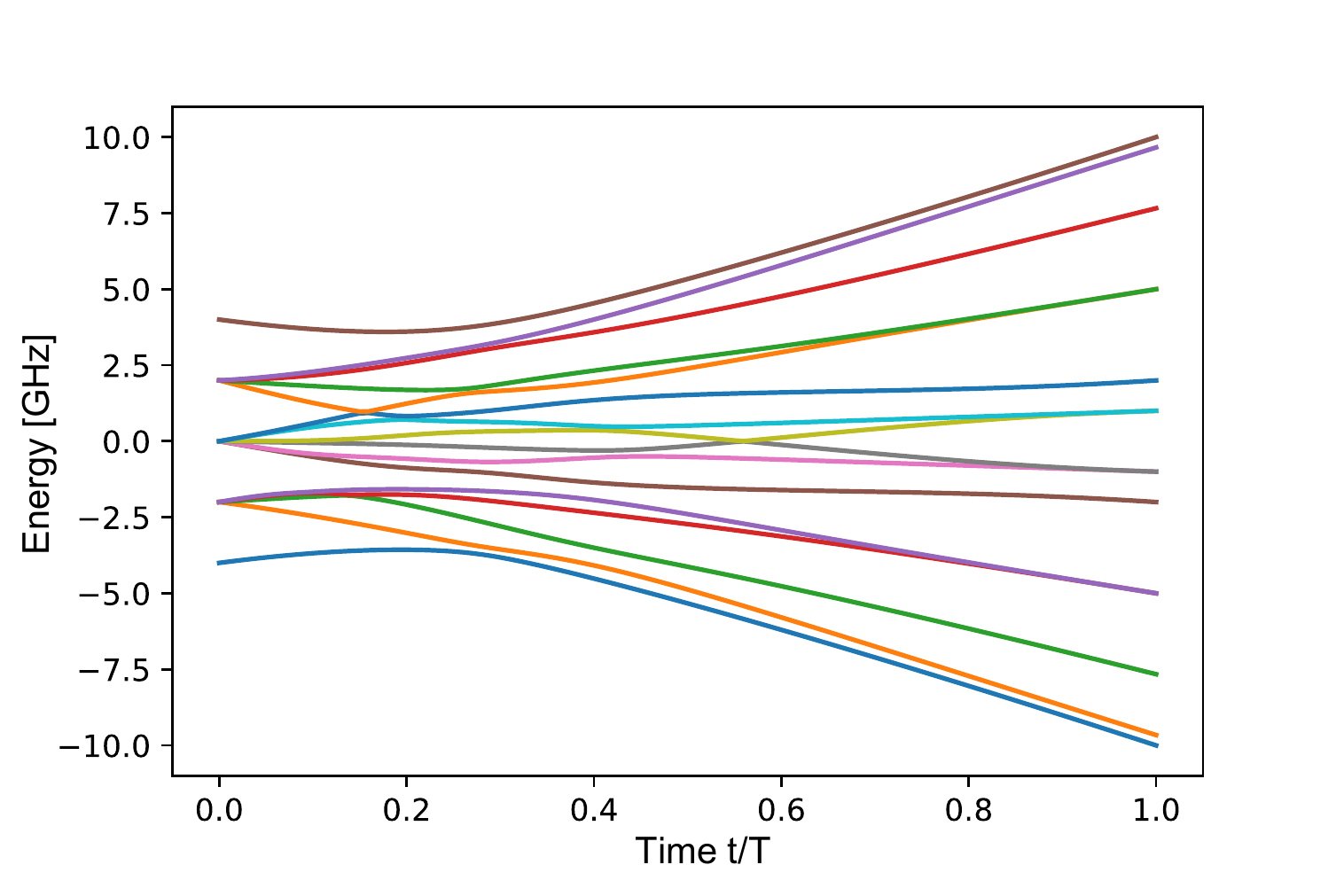}\\
(a)  Energy spectrum of QA for the deformed spin star model with the drive Hamiltonian of the transverse field.
\end{center}
\end{minipage}\\
\begin{minipage}{0.90\textwidth}
\begin{center}
\includegraphics[width=0.95\textwidth]{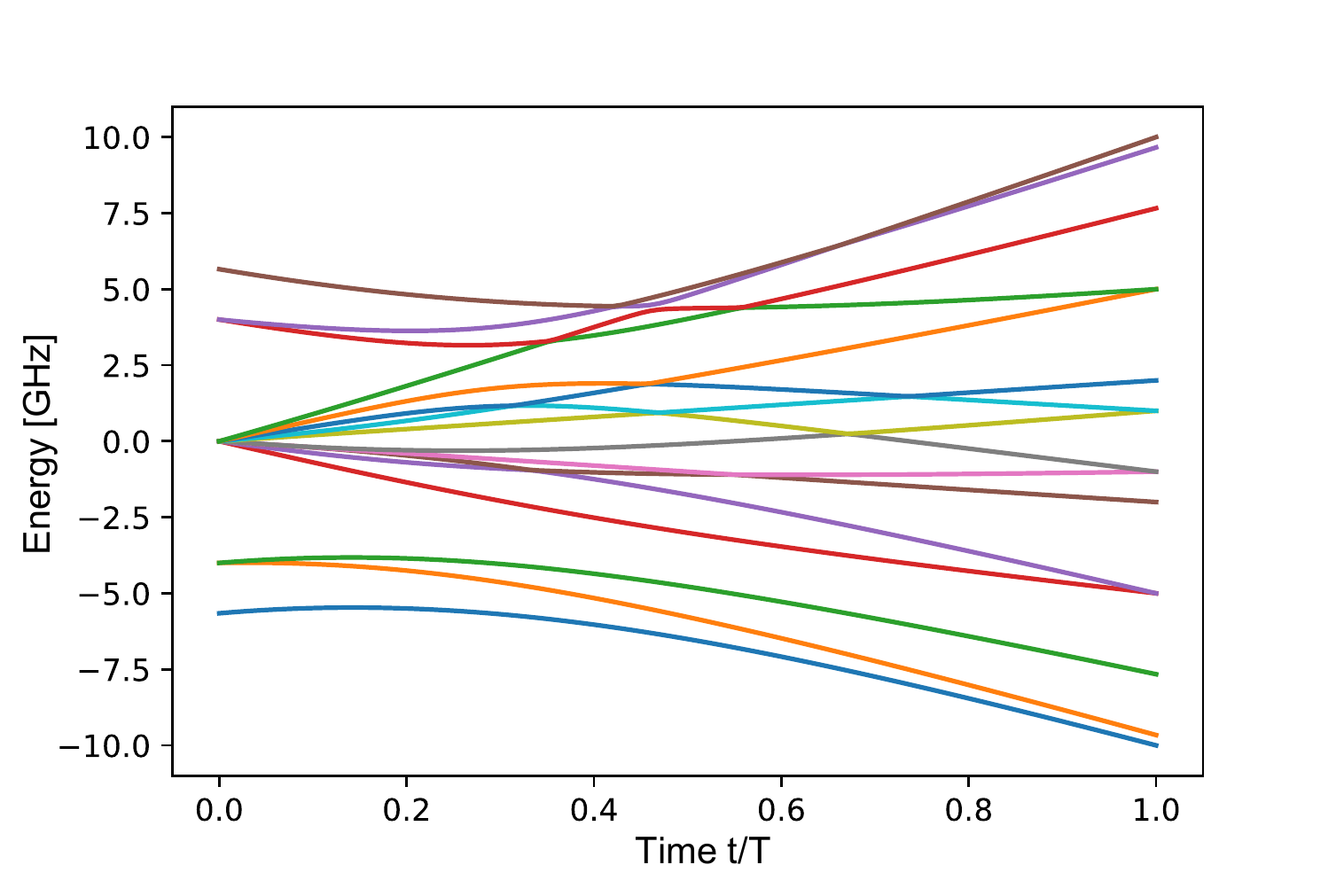}\\

(b)  Energy spectrum of QA for the deformed spin star model with the drive Hamiltonian of the XY model.
\end{center}
\end{minipage}
\end{center}
\end{minipage}
\caption{Energy spectrum during QA plotted against time $t$
to prepare the ground state of the deformed spin star model.
(a) The transverse field is chose.
(b) The XY model is chosen.
}
\label{fig:hydro_energy_spec}
\end{center}
\end{figure}

In this subsection, we consider the deformed spin star model as the problem Hamiltonian.
The Hamiltonian of the deformed spin star model is given by
\begin{align}
    H&= \omega\hat{\sigma}_{0}^{z}+\omega_{1}\hat{J}^{z}+J(\hat{\sigma}_{0}^{+}\hat{J}^{-}+\hat{\sigma}_{0}^{-}\hat{J}^{+})\label{eq:deformed_spin_star},
\end{align}
where $\hat{J}^{+}\equiv\sum_{j=1}^{L}e^{2\pi\frac{j}{L}}\sigma_{j}^{+}$ and $\hat{J}^{-}\equiv\sum_{j=1}^{L}e^{-2\pi\frac{j}{L}}\sigma_{j}^{-}$.
This kind of model has been investigated, because this describes a dynamics of a hybrid system composed of a superconducting flux qubit and nitrogen-vacancy centers in a diamond \cite{marcos2010coupling,twamley2010superconducting,zhu2011coherent,zhu2014observation,matsuzaki2015improving,cai2015analysis}.

\begin{figure}[ht]

\includegraphics[width=80mm]{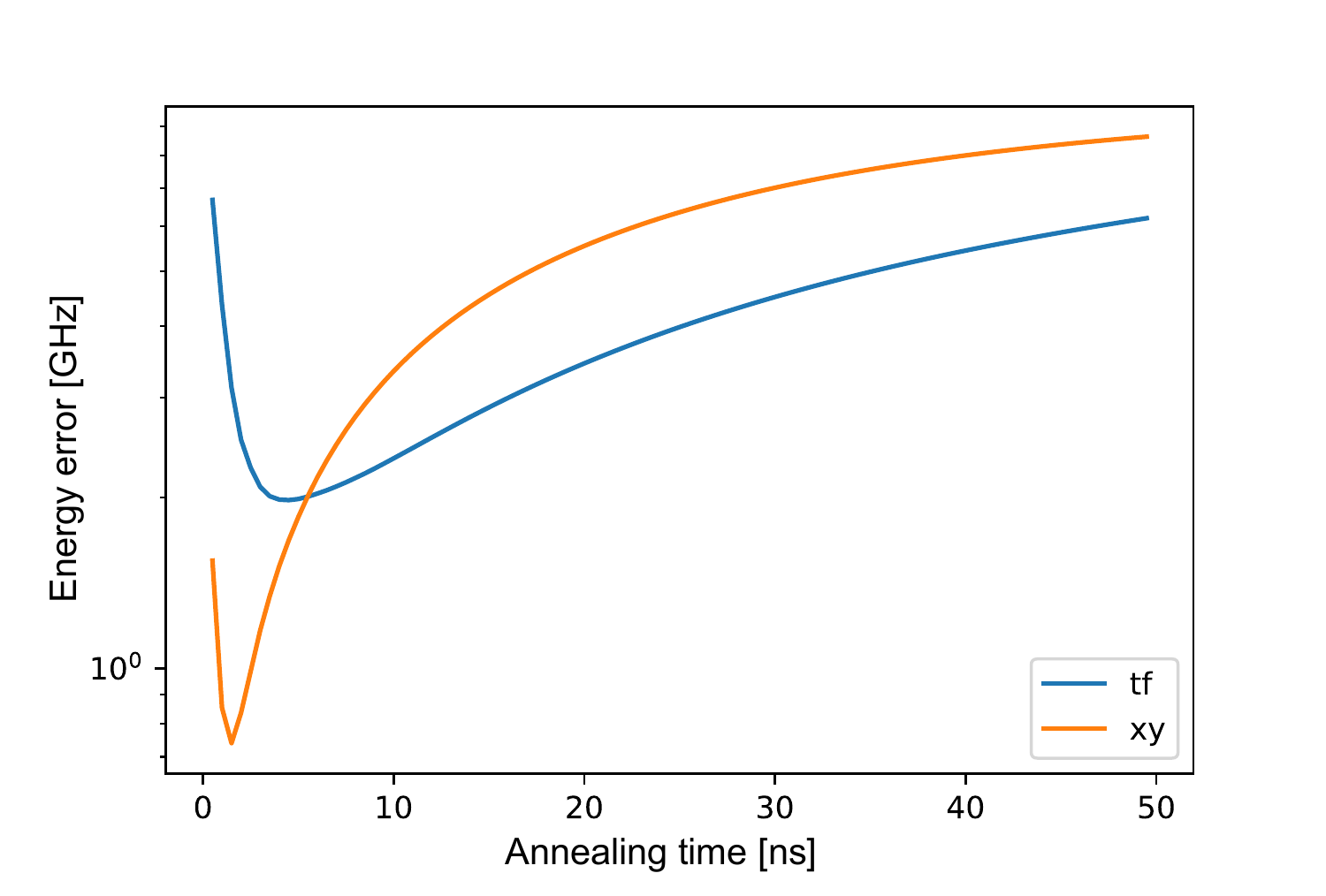}

\caption{
Estimation error $|E^{(\rm{true})}_g-E^{(\rm{QA})}_g|$ for the deformed spin star model plotted against the annealing time, where $E^{(\rm{true})}_g$ denotes the true ground state energy and $E^{(\rm{QA})}_g$ denotes the energy obtained by QA. 
The blue (orange) line denotes the case when we choose the XY model (transverse field) with an amplitude of $g$ ($B$) as the drive Hamiltonian. Here, for a optimal amplitude with a decoherence rate of $\gamma=2.5\times10^{-5}$ GHz.
In addition, we choose the model parameters $\omega=\omega_{1}=0.5$ GHz, $J=5$ GHz, and $L=3$(i.e. the total spin number is $4$).
}
\label{fig:hydro_const_log}

\end{figure}

The energy spectra of the annealing Hamiltonians for the conventional scheme and our scheme are plotted in Fig. \ref{fig:hydro_energy_spec}.
It is worth mentioning that the performance of QA depends on the amplitude of the drive Hamiltonian, i.e., $B$ and $g$. Hence, we sweep these parameters and compare the performance of our scheme with that of the conventional one when we optimize the amplitude of the drive Hamiltonian.
We performed the numerical simulations with decoherence rate of $\gamma=2.5\times10^{-5}$ GHz.
In the case, we choose the optimal amplitude of the drive Hamiltonian at each time.
We can see that the estimation error of our scheme is just three times smaller than that of the conventional scheme, as shown in Fig \ref{fig:hydro_const_log}.
We find that the optimal annealing time of our scheme, which provides the smallest estimation error, is much shorter than that of 
the other scheme.
This implies that, as the effect of the non-adiabatic transitions is significantly reduced, we can shorten the annealing time and thus suppress the effect of the decoherence.

\subsection{Random XXZ spin chain}

In this subsection, we consider a random XXZ spin chain as the problem Hamiltonian.
The Hamiltonian of the random XXZ spin chain is given by
\begin{align}
    H=\sum_{j=1}^{L-1}J_{j}\biggl(\sigma^{x}_{j}\sigma^{x}_{j+1}+\sigma^{y}_{j}\sigma^{y}_{j+1}+\Delta\sigma^{z}_{j}\sigma^{z}_{j+1}\biggr)\label{eq:xxz_spin_chain},
\end{align}
where $\Delta$ is an anisotropic parameter and $\{J_{j}\}_{j=1}^{L-1}$ is a random interaction.
We employ an open boundary condition, and we assume that $\{J_{j}\}_{j=1}^{L-1}$ is chosen from independent uniform distributions on the interval $[0,2]$ (i.e., $\{J_{j}\}_{j=1}^{L-1}\overset{\text\small\textrm{iid}}{\sim}U(0,2)$).
It is worth mentioning that this Hamiltonian commutes with the total magnetization $S_{z}$; hence, we can use our scheme to obtain the ground state of this Hamiltonian.


\begin{table}[htb]
\centering
    \begin{tabular}{|c||c|}
    \hline
      $J_{1}$ & $0.8441683664299817$ \\
      $J_{2}$ & $0.47574391516586223$ \\
      $J_{3}$ & $0.06980280523824778$ \\
      $J_{4}$ & $0.6197240483819366$ \\ \hline
    \end{tabular}
  \caption{Coupling constant of the random XXZ spin chain $\{J_{j}\}_{j=1}^{4}$ used in Fig. \ref{fig:random_xxz_energy_spec}. The unit of these values is GHz, as described in the main text.}
  \label{tb:rand_XXZ_coefficient_jw}
\end{table}

\begin{figure}[htbp]
\begin{center}
\begin{minipage}{0.48\textwidth}
\begin{center}
\begin{minipage}{0.90\textwidth}
\begin{center}
\includegraphics[width=0.95\textwidth]{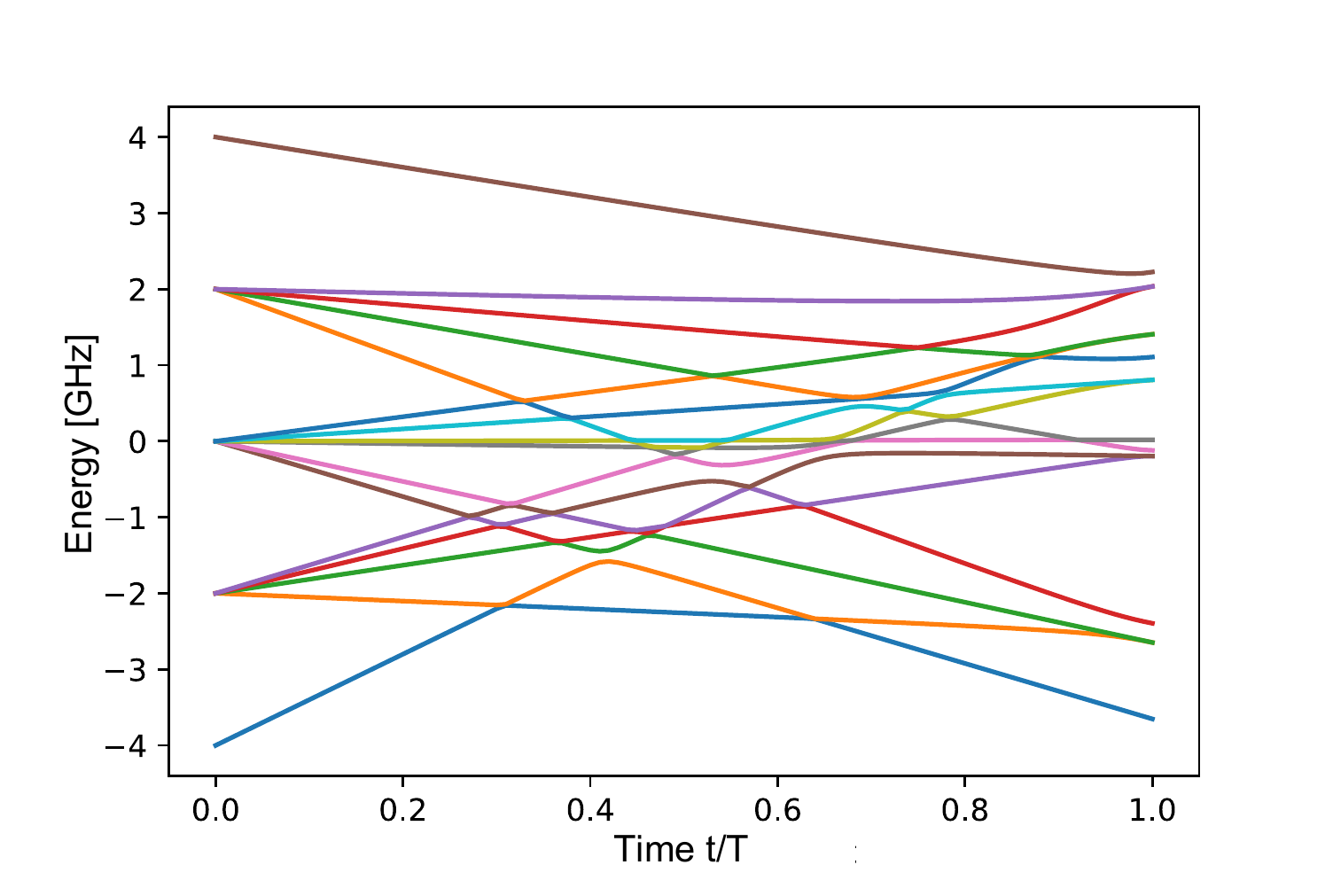}\\

(a) Energy spectrum of QA for the random XXZ spin chain with the drive Hamiltonian of tshe random transverse field.

\end{center}
\end{minipage}\\
\begin{minipage}{0.90\textwidth}
\begin{center}
\includegraphics[width=0.95\textwidth]{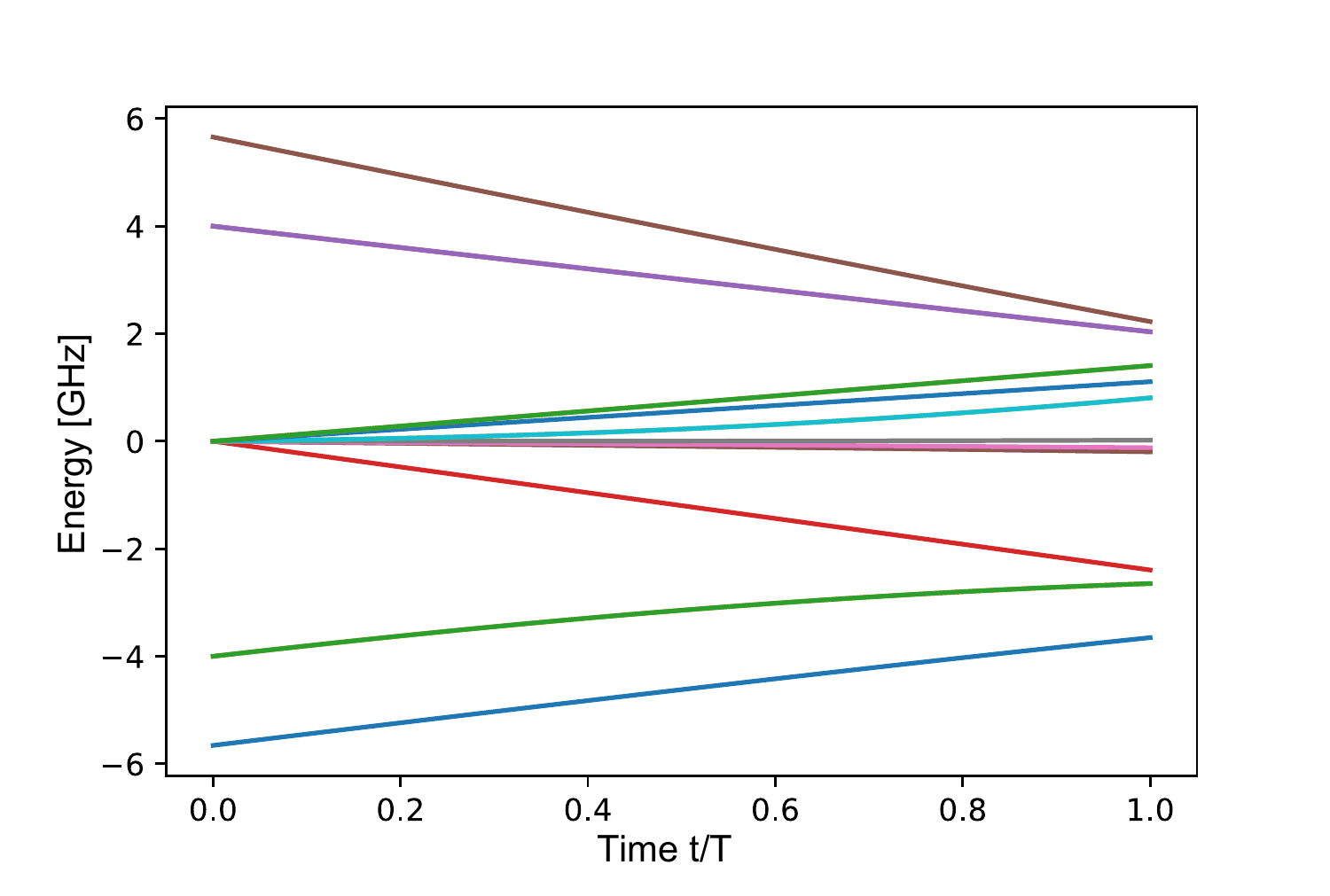}\\

(b) Energy spectrum of QA for the random XXZ spin chain with the drive Hamiltonian of the random XY model.
\end{center}
\end{minipage}
\end{center}
\end{minipage}

\caption{Energy spectrum during QA plotted against time to prepare the ground state of the random XXZ spin chain. The anisotropic parameter is
 $\Delta=0.7$ GHz and the number of qubits is $L=4$.
 The coupling strength of the XXZ spin chain is described in Table \ref{tb:rand_XXZ_coefficient_jw}.
(a)The  uniform transverse  field  is  chosen  as  the  drive Hamiltonian with an amplitude of $B=1.0$ GHz.
(b)The  XY model  is  chosen  as  the  drive Hamiltonian with an amplitude of $g=1.0$ GHz.
}
\label{fig:random_xxz_energy_spec}
\end{center}
\end{figure}

\begin{figure}[ht]

\includegraphics[width=80mm]{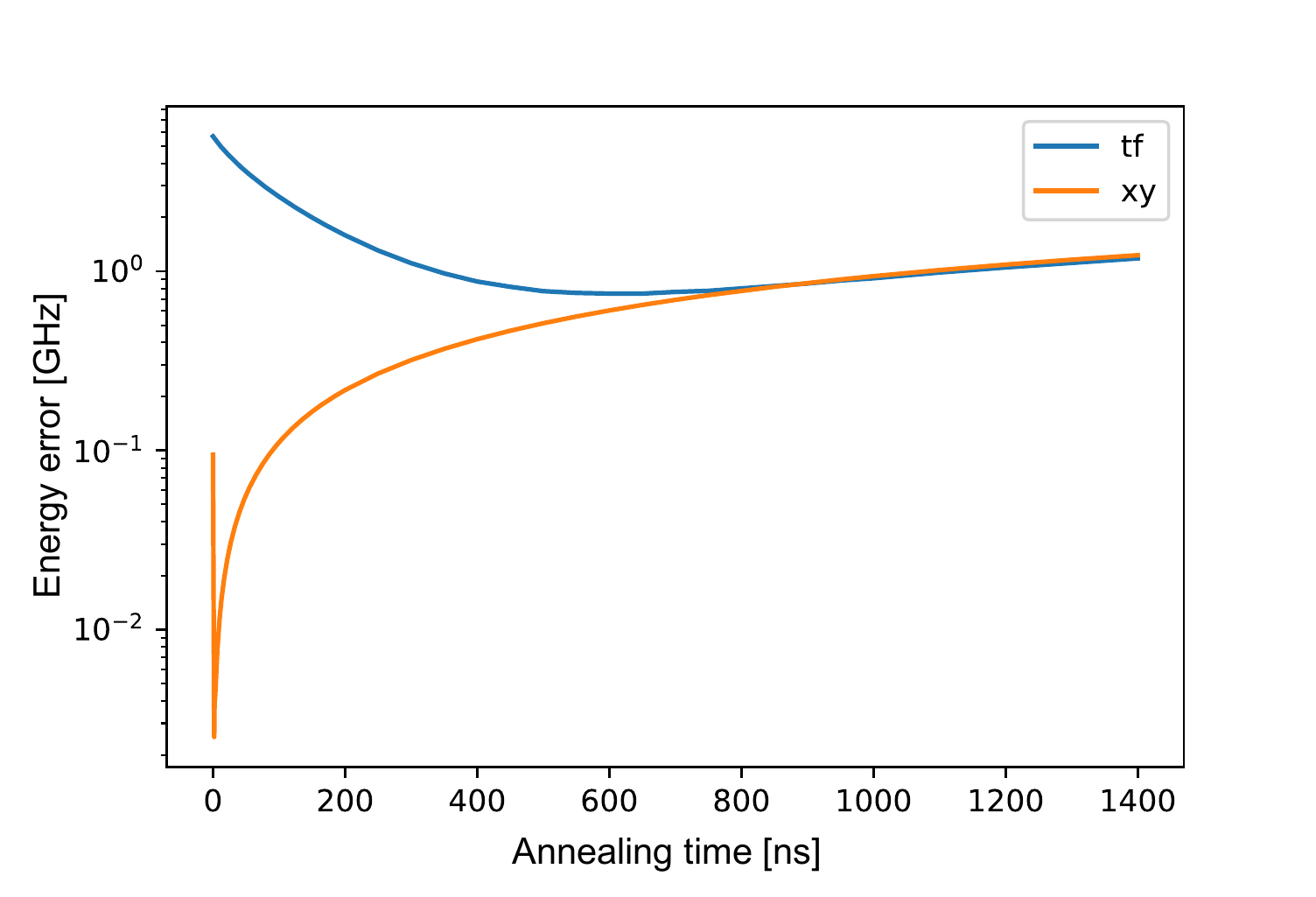}

\caption{Estimation error plotted against the annealing time. We choose the drive Hamiltonian as the uniform transverse field (blue) and the XY model (orange) with a decoherence rate of $\gamma=10^{-4}$ GHz. We use the same parameters as Fig. \ref{fig:random_xxz_energy_spec}.
}
\label{fig:random_xxz_random_tf}

\end{figure}

The energy spectrum of the annealing Hamiltonian for each scheme is plotted in Fig. \ref{fig:random_xxz_energy_spec}.
Furthermore, the estimation error is plotted against the annealing time in Fig. \ref{fig:random_xxz_random_tf}.
We performed numerical simulations with a decoherence rate of $\gamma=10^{-4}$ GHz.
In this case, we choose the optimal amplitude of the drive Hamiltonian at each time.
The estimation error of our scheme is around 10 times smaller than that of the other schemes.
Similar to the case of the deformed spin star model, we find that the optimal annealing time of our scheme becomes shorter than that of the conventional schemes, which confirms the suppression of the non-adiabatic transitions in our scheme.

\section{Conclusion}\label{sec:conclusion}

We proposed the use of a drive Hamiltonian that preserves the symmetry of the problem Hamiltonian. Owing to the symmetry, we can search the solution in an appropriate symmetric subspace during QA. As the non-adiabatic transitions occur only inside the specific subspace, our approach can potentially suppress unwanted non-adiabatic transitions. To evaluate the performance of our scheme, we employed the XY model as the drive Hamiltonian in order to find the ground state of problem Hamiltonians that commute with the total magnetization along the z-axis. We found that our scheme outperforms the conventional scheme in terms of the estimation error of QA.

\begin{acknowledgments}
We thank a useful comment from Shiro Kawabata.
This work was supported by MEXT's Leading Initiative for Excellent Young Researchers and JST PRESTO (Grant No. JPMJPR1919), Japan. This paper is partly
based on the results obtained from a project, JPNP16007,
commissioned by the New Energy and Industrial Technology Development Organization (NEDO), Japan.
\end{acknowledgments}

~~~


\bibliography{apssamp}

\end{document}